\documentstyle[aps,prl]{revtex}
\begin{document}
\title{ Gapless Singlet modes in the Kagom{\'e} strips: A study through DMRG and
strong coupling analysis}
\author{Swapan K. Pati and Rajiv R. P. Singh}
\address{Department of Physics, University of California, 
Davis, California 95616}

\twocolumn[\hsize\textwidth\columnwidth\hsize\csname
@twocolumnfalse\endcsname
\date{\today}
\maketitle
\widetext
\begin{abstract}
\begin{center}
    \parbox{6in} {Recently Azaria et al have studied strips of the 
Kagom{\'e}-lattice in the weak-coupling limit, where they consist of two 
spin-half chains on the outside weakly coupled to
an array of half-integer spins in the middle. Using a number of 
mappings they have
arrived at the interesting result that in this system all spin excitations
are gapped but there are gapless spinless modes. Here we study these
Kagom{\'e} strips in the limit where the interchain couplings are comparable
to the coupling to the middle spins by density matrix renormalization group
and by a strong coupling analysis. In the limit when the coupling 
to the middle-spin
dominates, the 5-spins of the unit-cell reduce to a single S=3/2 spin, and the
overall system has well known gapless spin excitations.
We study the phase transition from this phase to the weak-coupling phase.
We also carry out a strong coupling analysis
away from the S=3/2 limit, where the five-spin blocks have four degenerate
ground states with S=1/2, which can be thought of as two spin and 
two pseudospin degrees of
freedom. A numerical study of this strong coupling model also suggests
a finite spin-gap.}

\end{center}
\end{abstract}

]

\narrowtext

The spin-half Kagom{\'e}-lattice antiferromagnet has proven to be a very
fascinating system. Studies based on exact diagonalization of
finite systems\cite{ex1,ex2}and  series expansions\cite{series,hight}
strongly suggest that the system has a quantum disordered  ground state.
Finite size studies show a gap to spin excitations, but the
most fascinating aspect of these numerical studies is the existence
of a large number of spin-zero excitations below the lowest triplet
state. Their number appears to grow exponentially with the size of
the system\cite{lhu1,mila,lhu2}. The question of whether these spin-zero 
excitations are gapped or gapless and whether they form a well defined 
excitation mode of the system has not been resolved.

In this respect an interesting system was recently studied by 
Azaria et al\cite{azaria}.
They considered a 3-chain strip of the Kagom{\'e}-lattice shown in Fig 1.
This is a one-dimensional system with half-integral spin per unit cell,
and is thus subject to the Lieb-Schultz-Mattis (LSM) theorem\cite{lsm}. 
Most 
interestingly, Azaria et al find that the system has no broken 
symmetries, a spin-gap
but gapless spin-zero excitations. This is a rather unusual possibility,
but permitted by the LSM 
theorem. Azaria et al study the system perturbatively in the weak 
coupling limit,
where the Kagom{\'e} strip reduces to two spin-half chains on the outside 
with couplings
$J_{\parallel}$ on the chains,
coupled weakly with the array of middle spins, with couplings $J_{\perp}$. 
Using a Majorana fermion
representation for the low energy degrees of freedom on the spin-chains
and employing a number of mappings, they conclude that these Kagom{\'e}
strips have a spin-gap, which is exponentially small in 
$J_{\perp}/J_{\parallel}$,
and there are gapless spin-zero modes. The importance of these studies to
the Kagom{\'e}-strip limit $J_{\perp}=J_{\parallel}$, and furthermore to
the Kagom{\'e}-lattice antiferromagnets remains unclear. Given the large number
of approximate mappings, an independent numerical study of the model is
clearly desirable and is performed here.

The limit $\lambda=J_{\perp}/J_{\parallel}\to 0$ is particularly 
difficult for a numerical 
study as the spin-gap becomes exponentially small in that limit.  In the opposite 
limit, $\lambda\to\infty$, the five-spin unit cell of the system reduces 
to spin-3/2 and the system
reduces to a spin-3/2 chain. This system  is
well known to have gapless spin excitations. Thus there must be a 
phase transition in the model
as a function of $\lambda$. We study this system by the density 
matrix renormalization group(DMRG) and find that the phase transition 
occurs at $\lambda_c=1.20(2)$.
For $\lambda<\lambda_c$, the system has a spin-gap and gapless 
spin-zero modes. The Kagom{\'e} strips corresponding to $\lambda=1$ 
belong to the spin-gap phase.

We have also studied the system of weakly coupled 5-spin blocks, where
the ground state in the blocks are different from the $\lambda\to \infty$ limit.
For large $\lambda$, the
ground states of the block are the four states of a spin-3/2 spin.
However at $\lambda=2$, there is a level crossing in the 5-spin blocks and for
$1<\lambda<2$, a new set of fourfold degenerate states become ground states.
These states have spin-half. They correspond to a singlet pair on
one of the chains and a triplet pair on the other chain, which is
combined with the middle spin into a spin-half. Thus in addition to spin
there is an additional two-fold degeneracy in every block that corresponds to an Ising
like degree of freedom.
We generate an effective Hamiltonian for the kagome strips by treating the
coupling between blocks in degenerate perturbation theory. A numerical
study of this strong coupling Hamiltonian also suggests the existence
of a spin-gap. Whether this
strong coupling phase is the same as the weak-coupling phase studied
by Azaria et al remains to be seen.

The Hamiltonian corresponding to the Kagom{\'e} strip can be written as
\begin{equation}
 H= \sum_{a=1,2} \sum_i [J_{\|}(S_{a,2i} S_{a,2i+1})+J_{\bot}
S_{2i+1/2} (S_{a,i}+S_{a,2i+1})]
\end{equation}
\noindent where there are two chains with index $a=1,2$ and the $S_{2i+1/2}$
is the middle spin which interacts with the four sites, two from 
each chain (see Fig.1).

 We have carried out an
 extensive DMRG \cite{white,dmrg} 
calculation on this Kagom{\'e} strip Hamiltonian, for various parameter
 ranges as well as for various system sizes with periodic boundary 
 conditions. Apart from conserving the z-component of the total spin
($S_{tot}^z$), we have also used spin-parity (up-down symmetry)
 to divide the total spin spectrum into even and odd total
 spin subspaces for accurate description of the states in singlet and triplet
 branches.

 The unit cell contains 5 spin-half sites, so it will be error-prone
to put one or two unit cells in every DMRG iteration. 
 Instead, we have
 increased the system size by 2 spin-half sites in every iteration.
 However, this creates new difficulties if we introduce the sites in the
conventional way. At intermediate stages of DMRG
 we need to introduce couplings between sites which
 are not coupled in the true system. This problem is reduced as far as
 possible by inserting
 sites in a particular order, closely resembling the true system. Furthermore, 
 the intermediate stage configurations and couplings, for every value
 of $\lambda$, are optimized by comparing the DMRG results for smaller 
systems with exact diagonalization.
 As accuracies in the energy are very important, we have taken extra
 precautions to check energies at every intermediate DMRG step.
 The DMRG accuracy depends on $m$, the number of density matrix eigenstates
 kept per block as well as the ratio $\lambda$. We use $m$ between
 $100$ to $200$. In the spin-3/2 (large $\lambda$) phase small values of $m$ 
 suffice, while large values of $m$ are necessary for the calculation of
 properties in the spin-gap phase. Finite
 size DMRG algorithm has been implemented for even numbers of 5-spin blocks,
 and depending on the energy convergence we have performed 2 to 3 
finite-system sweeps. Truncation errors, defined by the sum of the 
discarded density matrix eigenvalues, ranged from zero to $10^{-6}$ for
$\lambda=1$. As has been discussed in various DMRG studies, this
discarded density matrix weight roughly measures the absolute error in the
DMRG energies. The largest system that we have studied varies from $50$ to
$100$ spin-half sites. We have set $J_{\bot}=1$ and varied the coupling, 
$J_{\|}$, from $0$ to $2$.

We begin with the DMRG
results on the ground state properties.
The ground state of the system
is a non-degenerate spin singlet over the entire parameter range. 
The ground state energy is a continuous function of $J_{\|}/J_{\bot}$.
However, its 2nd derivative 
with respect to the $J_{\|}/J_{\bot}$, has a kink at
$J_{\|}/J_{\bot}=0.8 \pm 0.05$, signalling a continuous phase transition.
To study this transition further,
we plot nearest neighbour correlation 
function between the spins in a chain, within a unit cell, as a function of
$J_{\|}/J_{\bot}$ in Fig.2. This correlation is positive when the 
system is an effective spin-3/2 chain, and the outer spins in a block
are all parallel. When the chain couplings dominate, all nearest neighbor
chain correlations become antiferromagnetic. The change in sign occurs precisely
at this transition.
Let us call the phase for 
$J_{\|}/J_{\bot} < 0.8 \pm 0.05$ phase I and  the phase for $J_{\|}/J_{\bot}$ 
$> 0.8 \pm 0.05$ phase II.

We have calculated various equal-time two spin correlation functions for the system.
Let us distinguish between four different types of pair correlations, namely,
those (i) between the spins in the middle row, (ii) between the middle 
spins and the spins in the outer chains, (iii) between the spins on the 
same chain and (iv) between spins on different chains.
In the DMRG procedure, we have
computed these correlation functions from the sites inserted at the last
iteration or it's previous iteration, to minimize the errors. Two such
correlations (type (i) and (iii)) are shown in Fig.3 in the two phases 
noted above. At small $J_{\|}/J_{\bot}$ (in phase I), both the
correlations decay algebraically with distance, 
which is consistent with the quasi long range order of the half-odd-integer
spin chains. But as we go to phase II, the correlation functions behave
differently. In this case, we observe a rapid decay of the
correlations. For the middle spin correlations, 
the decay is quite fast and eventually settles down
to a number of order $10^{-3}$, which within our numerical
uncertainties is consistent with zero.
Thus the behavior is suggestive of a short correlation length in phase II.

We now present results on the excitation spectrum.
The spin-gap can be defined as the energy difference between the lowest
energy states in even and odd parity branches, which is equivalent to the
lowest energies in $S^z_{tot}=0$ and $S^z_{tot}=1$ sectors respectively. 
We have targetted lowest few states in the even parity branch and ensured
that these are singlet states as they do not appear in $S^z_{tot}=2$ sector.
In Fig.4, we show the excitation gaps to the lowest few excited states
from the ground state singlet in phase II. As is clear from the figure, 
there is a finite gap to the lowest triplet state, but not to the singlets. 
If we had a system with broken symmetry, as in the Majumdar-Ghosh model,
we would expect two singlet states to become degenerate in the thermodynamic
limit, but a gap to all other states\cite{mgmodel,wh-aff}. This is not the 
case here. We see
that a number of singlets are coming down in energy as the system size
is increased, suggesting gapless singlet modes. As we have not done a 
wavevector resolved calculation, we cannot tell how many gapless singlet
modes are present. Our calculations are also not accurate enough to
determine the central charge of the system in phase II, which will also
give the number of gapless modes.

We have observed a finite spin-gap all the way from 
$J_{\parallel}/J_{\perp}=0.85 \pm 0.05$ to $2.0$. However, the spin-gap 
is quite small over the entire parameter range and the maximum in the 
spin-gap occurs at around
$J_{\parallel}/J_{\perp}=1.1$, with a gap value of $0.15 \pm 0.01$. This spin-gap
measurement is quite reminiscent of what White and Affleck had found in the
zig-zag chain at large next nearest neighbour coupling\cite{wh-aff}. It is
difficult to calculate the small gaps accurately. A better 
alternative was
to find the correlation length (instead of gap) by DMRG. Unfortunately,
because of our large unit cell, we are not able to go to large enough systems
to calculate the correlation length.
The spin gap would eventually vanish as $J_{\parallel}/J_{\perp}$ becomes 
large as in that limit there are two isolated spin-1/2 chains and
free spin-1/2's in the middle. 

We now turn to a strong-coupling analysis of the Kagom{\'e}-strip problem.
By the strong coupling limit, we mean a limit in which the Kagom{\'e}-strip
can be regarded as weakly coupled 5-spin blocks. As discussed earlier,
the spin-3/2 phase is clearly well described in the strong-coupling limit.
The question we wish to address
is could the spin-gap phase be desribed by a strong-coupling approach.
To this end, we first let the inter-block
exchange be zero and vary $\lambda=J_{\perp}/J_{\parallel}$ within a block
and examine the ground state multiplets. For large $\lambda$, we have a
spin-3/2 ground state, which turning on interblock interactions leads to
the spin-3/2 chain. At $\lambda=2$, there is a level crossing transition
and for $1<\lambda<2$ a different set of four-fold degenerate states
becomes the ground state for the block. A potential candidate for the spin-gap
phase is one where these four states of a block are coupled by the
interblock interactions.  Note that these four states have spin-half
and consist of singlet on one of the chains and a triplet on the
other chain, which is then combined into a spin-half with the
spin in the middle. Thus in addition to the spin degree of freedom,
there is an additional spin-half `pseudospin' degree of freedom.
The pseudospins have Ising-like symmetry which corresponds to interchanging
the two chains. 

We can now use degenerate perturbation theory to obtain
an effective Hamiltonian coupling the spin-pseudospin degrees of freedom
on neighboring blocks. Before proceeding with the degenerate perturbation
theory, we note that at $\lambda=1$, these $4$-states become degenerate
with two other states of the block. These are states, where in both chains
the spins are combined into a singlet and the middle spin is free. We note
that an extension of Mila's proposed low energy manifold for the 
Kagom{\'e} lattice\cite{mila}, will
reduce to these 6-states per block for the Kagom{\'e} strip. However, for now,
we will restrict ourselves to $4$-states only and leave the $6$-state
degenerate perturbation theory for later work.

To proceed, we decompose the full Kagom{\'e}-strip Hamiltonian 
as $H=H_0+ H_1$, where $H_0$ contains parts of the
intra-block interactions and $H_1$ contains rest of intra-block and all of
the interactions between the blocks. To be concrete, we consider,
\begin{eqnarray}
H_0 =S_{2i+1/2}(S_{1,2i}+S_{2,2i}+S_{1,2i+1}+S_{2,2i+1}) \nonumber \\
+3/4(S_{1,2i}S_{1,2i+1}+S_{2,2i}S_{2,2i+1})
\end{eqnarray}
\noindent and
\begin{eqnarray}
H_1=J_{\|}(S_{1,2i+1}S_{1',2i+1} +S_{2,2i+1}S_{2',2i+1}) \nonumber \\
+1/4(S_{1,2i}S_{1,2i+1}+S_{2,2i}S_{2,2i+1})
\end{eqnarray}
\noindent where $S_{k,2i}$ denotes the spin at site $2i$
of the k-th chain. And the chain index with a $\it prime$
indicates that the chain belongs to another block.

The first order effective
Hamiltonian turns out to be quite simple and without quantum 
fluctuation in the pseudospin degrees of freedom.
\begin{equation}
H_{eff}^1 = J_{\|}[-1/4 + 2/9\sum_{ij}(S_iS_j + 4S_iS_jT_i^zT_j^z)]
\end{equation}
\noindent where $S_i$ and $T_i$ are the spin and pseudospin operators
respectively. The second order effective
Hamiltonian is given by,
\begin{eqnarray}
H_{eff}^2  = & & J_{\|}^2\sum_{ij}[-0.5817+0.1201S_iS_j+0.3041T_i^{z}T_j^{z} 
\nonumber \\
& & -0.1126(T_i^{+}T_j^{-}+T_i^{-}T_j^{+})-0.119(T_i^{+}T_j^{+}+T_i^{-}T_j^{-})
\nonumber \\ 
& & +2.2428S_iS_jT_i^{z}T_j^{z} -0.2334S_iS_j(T_i^{+}T_j^{-}+T_i^{-}T_j^{+})
\nonumber \\
& & +0.2344S_iS_j(T_i^{+}T_j^{+} +T_i^{-}T_j^{-})]
\end{eqnarray}
Here, the biquadratic exchanges, which involve four operator terms, 
could not be made simple. We checked the validity of this
perturbation theory by making sure that the Hamiltonian
written above produces the correct low-energy spectrum 
for a two-block system of the 
original Hamiltonian, $H=H_0+c H_1$, with small enough $c$.

We now consider a system described by a Hamiltonian
$H_{eff}^1+H_{eff}^2$. In other words, we are considering the perturbations
to be of order one but neglecting higher order terms.
This Hamiltonian has $4$-states per site and can be studied by
numerical methods.
We use an exact diagonalization 
method to study the system with up to 10 sites.

The ground state with $2N$ sites has $S^z_{tot}=0$. Although the sizes
studied are small, it strongly suggests a gapped triplet excitation
as $N \to \infty$. Our data are not
conclusive enough to identify whether the phase corresponds to a broken
symmetry ground state or the Kagom{\'e} phase with gapless nonmagnetic
excitations. Note that this Hamiltonian has similarities to biquadratic
spin-orbital model which has a broken symmetry spin-gap phase\cite{so1,so2,so3}.
Thus, the question remains open whether the phase II of the Kagom{\'e} 
strip can be obtained within a strong coupling approach.

In conclusion, in this paper we have studied the 3-chain Kagom{\'e} strip 
Hamiltonian by
DMRG and by a strong coupling expansion. The results support
the work of Azaria et al that in this model there is a spin-gap phase
with gapless singlet modes.

Acknowledgements: We thank Diptiman Sen and S. Ramasesha for valuable 
discussions. This work was supported by a grant from NSF 
number DMR-96-16574.

{\bf Figure Captions:}

\vspace*{0.5cm}

Fig.1: Schematic diagram of the Kagom{\'e} strip lattice. The corresponding
Hamiltonian is given in Eq.~1.

\vspace*{0.3cm}

Fig.2: Nearest neighbour two-spin correlation functions between the
spins on the chain within a unit cell as a function of $\lambda$, the
coupling ratio.

\vspace*{0.3cm}

Fig.3: Two-spin correlation functions a) between the spins on an outer chain
and b) between the middle spins,
in the two phases.

\vspace*{0.3cm}

Fig.4: Excitation gaps as a function of system size. The lowest triplet excitation
and three lowest singlet excitations are shown.

\begin{references}
\bibitem{ex1} C. Zeng and V. Elser, \prb, {\bf 42}, 8436 (1990).
\bibitem{ex2} J. T. Chalker and J. F. Eastmond, \prb, {\bf 46}, 14201 (1992);
P. W. Leung and V. Elser, \prb, {\bf 47}, 5459 (1993).
\bibitem{series} R. R. P. Singh and D. A. Huse, \prl, {\bf 68}, 1766 (1992).
\bibitem{hight} N. Elstner and A. P. Young, \prb, {\bf 50}, 6871 (1994).
\bibitem{lhu1} P. Lecheminant, B. Bermu, C. Lhuillier, L. Pierre and P. Sindzingre, \prb, {\bf 56}, 2521 (1997).
\bibitem{mila} F. Mila, \prl, {\bf 81}, 2356 (1998).
\bibitem{lhu2}C. Waldtmann, H.-U. Everts, B. Bernu, P. Sindzingre, 
C. Lhuillier, P. Lecheminant and L. Pierre, Euro. Phys. J. B., {\bf 2},
501 (1998).
\bibitem{azaria} P. Azaria, C. Hooley, P. Lechheminant, C. Lhuillier and
A. M. Tsvelik, \prl, {\bf 81}, 1694 (1998).
\bibitem{lsm} E. Lieb, T. Schultz and D. Mattis, Ann. Phys. (N.Y.),
{\bf 16}, 407 (1961).
\bibitem{white} S. R. White, \prl, {\bf 69}, 2863 (1992); \prb, {\bf 48},
10345 (1993).
\bibitem{dmrg} S. R. White and D. A. Huse, Phys. Rev. B, {\bf 48}, 3844 
(1993); Y. Kato and A. Tanaka, J. Phys. Soc. Jpn., {\bf 63}, 1277 (1994); 
S. K. Pati, S. Ramasesha and D. Sen, Phys. Rev. B,
{\bf 55}, 8894 (1997); J. Phys. Condens. Matt., {\bf 9}, 8707 (1997).
\bibitem{mgmodel}C. K. Majumdar and D. P. Ghosh, J. Math. Phys., {\bf 10},
1388, 1399 (1969)
\bibitem{wh-aff} S. R. White and I. Affleck, \prb, {\bf 54}, 9862 (1996).
\bibitem{so1} A. K. Kolezhuk and H. J. Mikeska, \prl,
{\bf 80}, 2709 (1998); Int. J. Mod. Phys. B, {\bf 12}, 2325 (1998).
\bibitem{so2} S. K. Pati, R. R. P. Singh and D. Khomskii, \prl,
{\bf 81}, 5406 (1998).
\bibitem{so3} P. Azaria, A. O. Gogolin, P. Lecheminant and A. A. Nersesyan,
preprint, cond-mat/9903047.
\end{references}
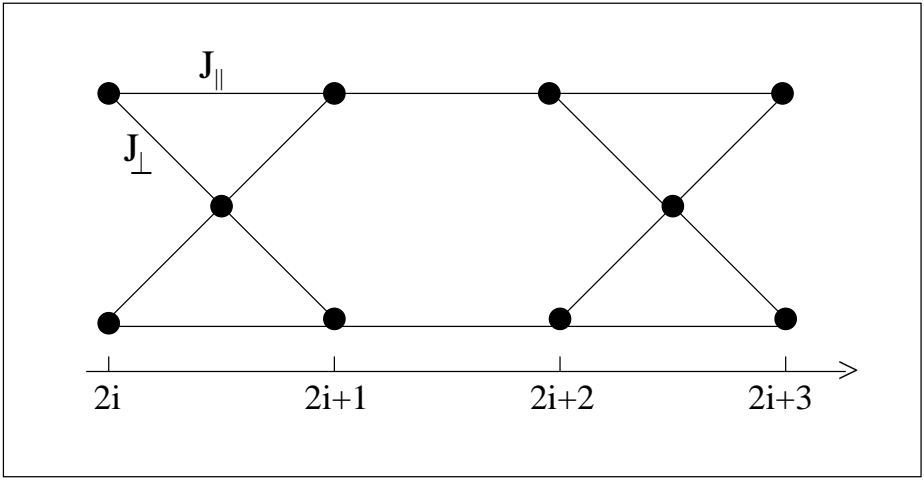
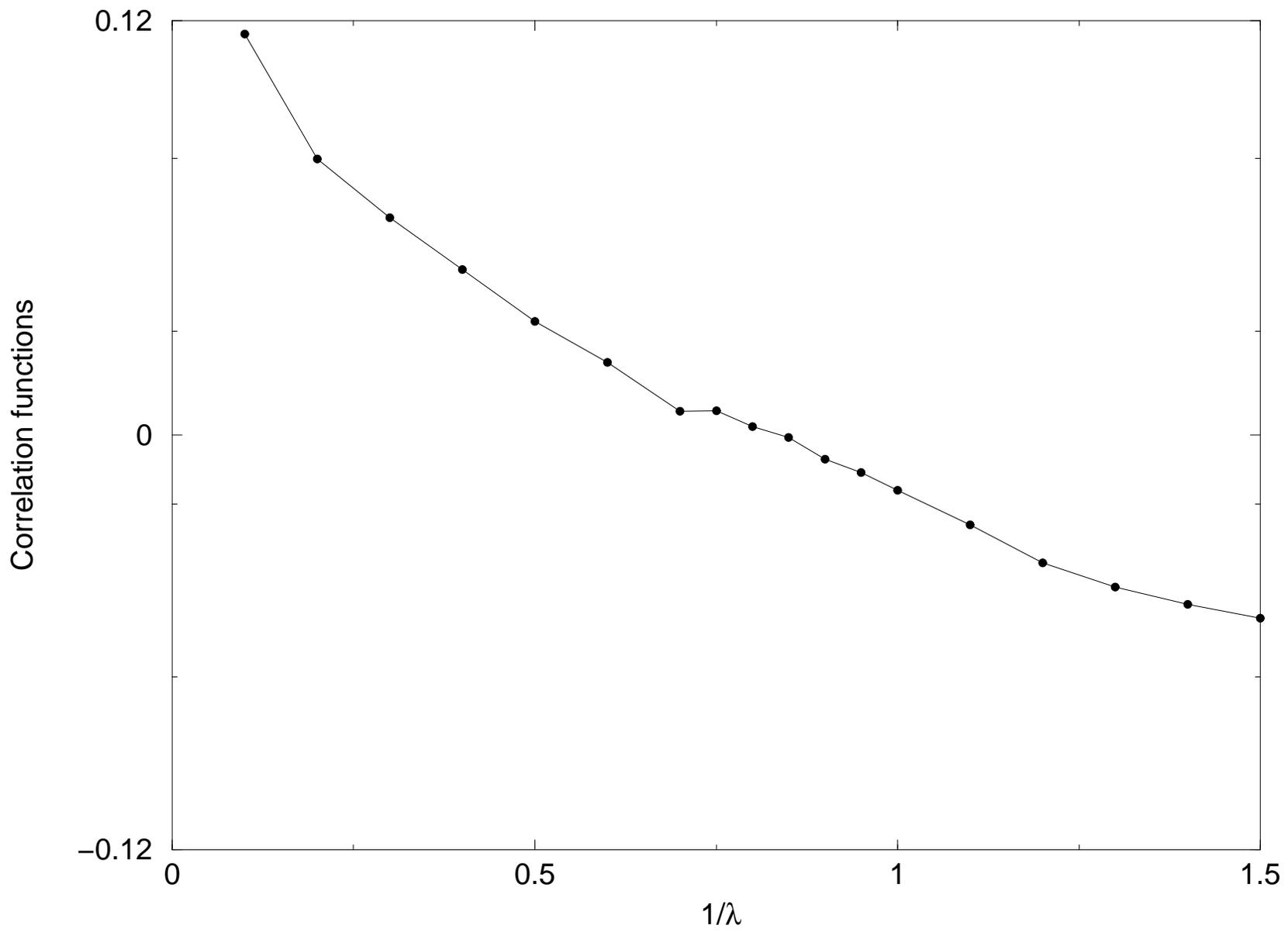
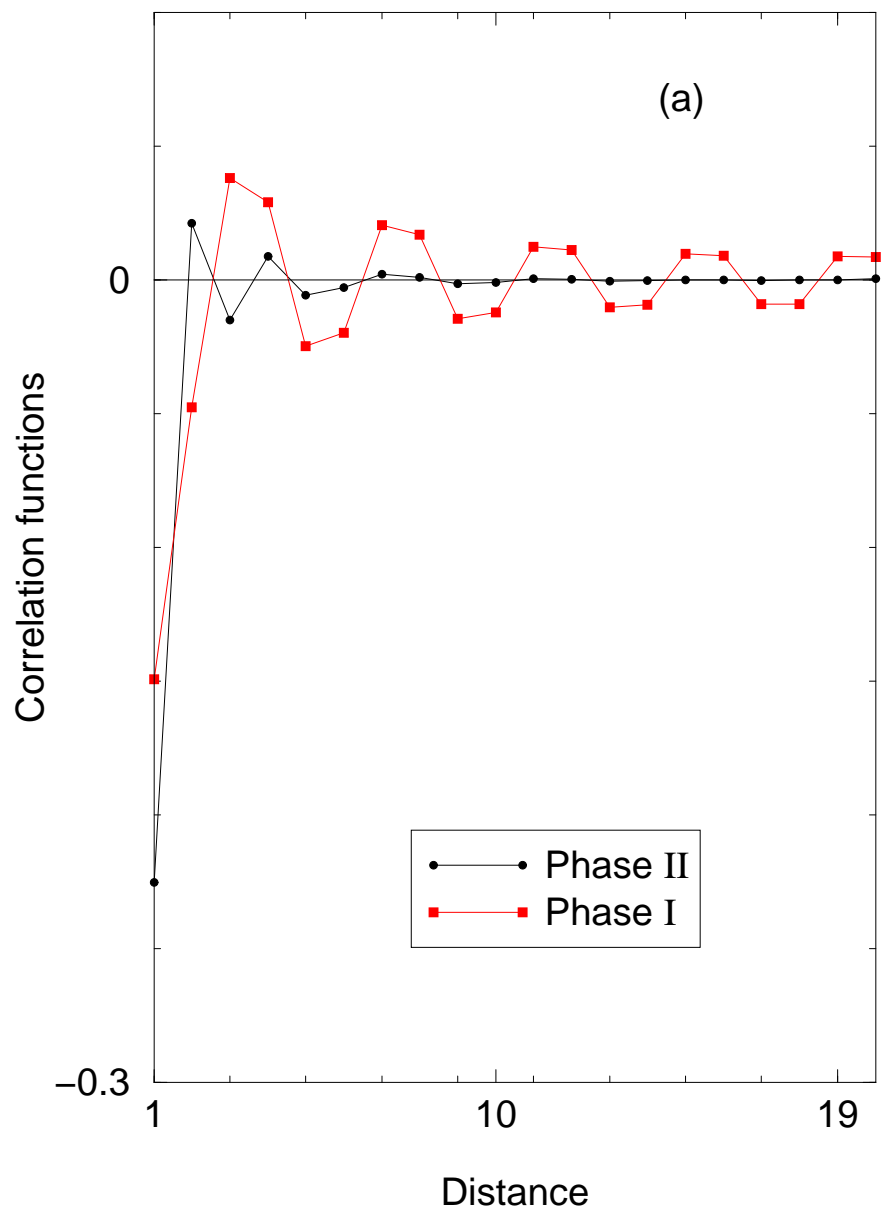
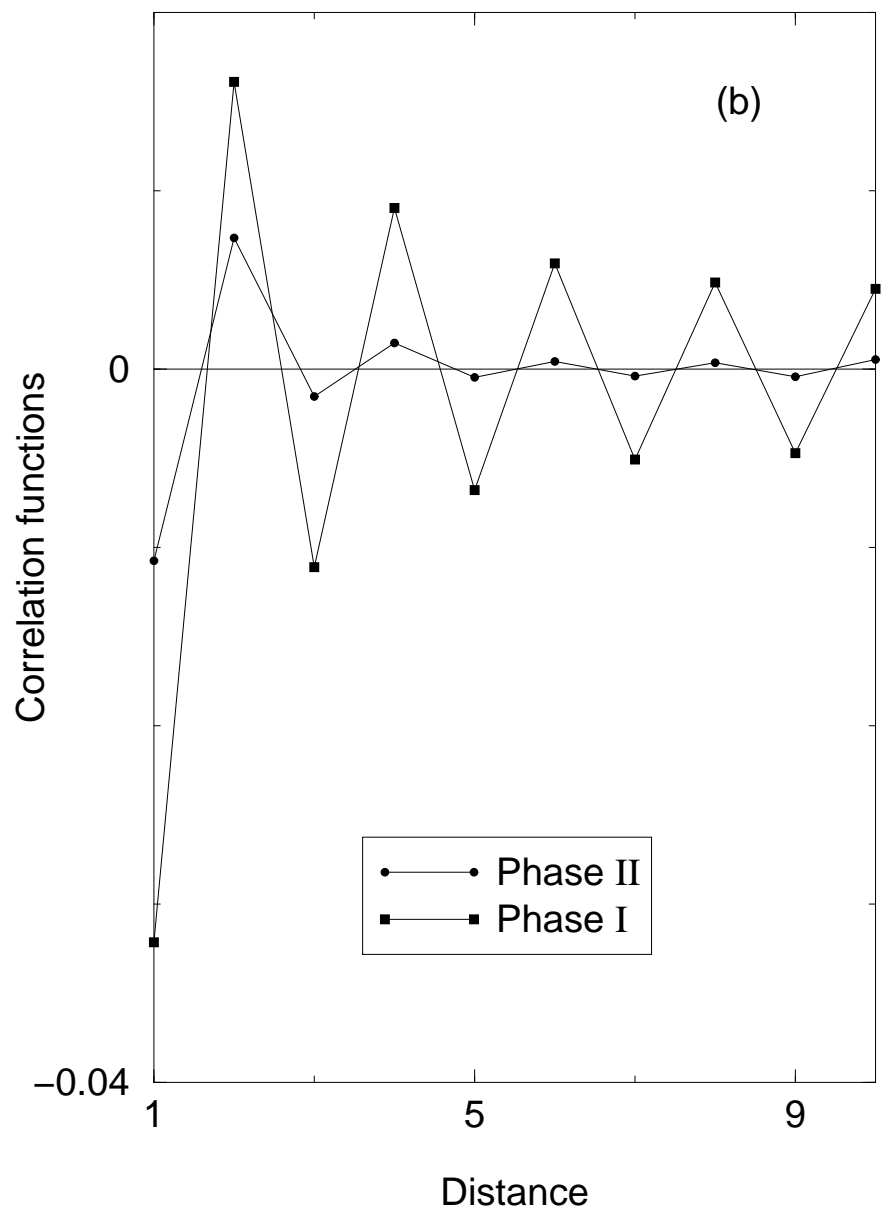
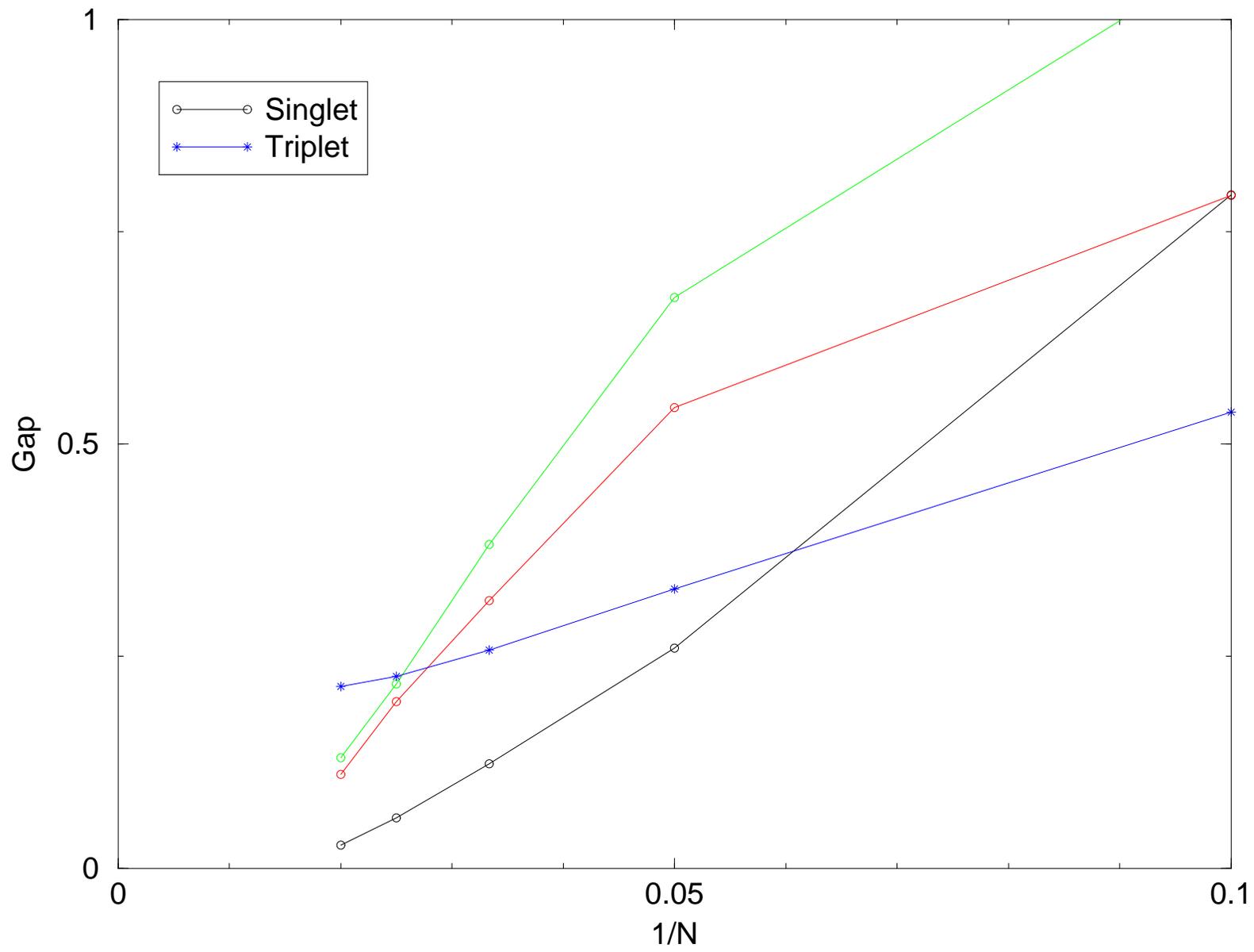
\end{document}